\documentclass[10pt]{revtex4}
\usepackage{amssymb}
\usepackage{latexsym}
\usepackage{epsfig}
\usepackage{float}
\usepackage{graphicx,epsfig, color }
\usepackage{graphicx}
\usepackage{subfigure}
\usepackage{amsmath}

\begin{document}
\title{Third order quasi-topological black hole with power-law Maxwell nonlinear source}
\author{ M. Ghanaatian$^{1}$\footnote{Corresponding author}, F. Naeimipour$^{1}$, A. Bazrafshan$^{2}$, M. Eftekharian$^{1}$}
\address{$^1$ Department of Physics, Payame Noor University (PNU), P.O. Box 19395-3697 Tehran, Iran\\
$^2$ Department of Physics, Jahrom University, 74137-66171 Jahrom, Iran}

\begin{abstract}
In this paper, we construct a new class of solutions for five dimensional third order quasi-topological black holes coupled to a power-law Maxwell nonlinear electrodynamics. To have real solutions, we should establish condition $\mu<-\frac{\lambda^2}{3}$ and to have finite solutions at infinity, the parameter of power-law Maxwell theory "s" should obey $\frac{1}{2}<s\leq 2$. Power-law Maxwell lagrangian is successful to set conformal invariance in higher dimensions. Also, this theory can reduce the divergence of the electrical field at the origin that is caused in linear Maxwell theory. As the value of parameter "s" increases, this divergence reduces more. In asymptotically anti-de sitter spacetimes, these obtained solutions lead to a black hole with two horizons for small values of $s$ and $q$. Also, solutions for $s=2$ have different behaviors with respect to the ones for other values of $s$. For negative and small values of parameter $\mu$, these solutions can describe a black hole with two horizons. An other important tip is that this black hole has thermal stability just for anti-de sitter solutions if its temperature is positive.

\end{abstract}


\maketitle

\section{Introduction}
Nonlinear electrodynamics was introduced with the motivation of solving some problems. First, this theory can eliminate or reduce the singularity of the electrical field of point charge at the origin in linear Maxwell theory. Second, some physical
systems in nature that contain field equations such as the ones in gravitational systems, are
intrinsically nonlinear. So, a proper action which leads to these field equations, should include nonlinear electrodynamics. 
Third, on the side of AdS/CFT correspondence, we can investigate the effects of nonlinear electrodynamic
gauge fields on the dynamics of strongly coupled dual theory \cite{Ge1,cai1}. Fourth, nonlinear electrodynamics has good physical properties about wave propagation,
such as the absence of shock waves and birefringence \cite{Boillat1}. Therefore, these reasons are enough for us to go to the nonlinear electrodynamics theory.\\ 
Three nonlinear lagrangians were introduced to solve the above problems: Born Infeld \cite{Born1}, logarithmic \cite{soleng} and exponential \cite{Hendi1} lagrangians. Born-Infeld and logarithmic lagrangians were successful to remove the divergency of the electrical field at the origin and so, they produce a finite value for the electrical field in $r=0$ \cite{Dey1,Hendi22}. However, the eletrical field of the origin diverges in the presence of Exponential lagrangian but its divergence is much slower than the ones in the presence of linear
Maxwell field \cite{Hendi22}.\\  
Later, an other new nonlinear lagrangian in the name of power-law Maxwell lagrangian was introduced. In addition to the above mentioned benefits, power-law Maxwell lagrangian has one more priority to the first three lagrangians. It can establish conformal invariance in higher dimensions. Under conformal transformations, $g_{\mu\nu}\rightarrow \Omega^2 g_{\mu\nu}$ and $A_{\mu}\rightarrow A_{\mu}$, the four dimensional black hole with Maxwell theory is invariant. In higher dimensions, there is a lack of conformal symmetry with Maxwell theory \cite{Tangher}. To solve this problem, we generalize to the $(n+1)$-dimensional action 
\begin{eqnarray}
S_{PM}=-\int_{\mathcal{M}} d^{n+1}x \sqrt{-g}[F_{\mu\nu}F^{\mu\nu}]^{s},
\end{eqnarray} 
where the energy-momentum tensor for this action is \cite{Hassaïne} 
\begin{eqnarray}
T_{\mu\nu}=4\bigg(sF_{\mu\rho}F_{\nu}^{\rho}(F_{\alpha\beta}F^{\alpha\beta})^{s-1}-\frac{1}{4}g_{\mu\nu}(F_{\alpha\beta}F^{\alpha\beta})^{s})\bigg).
\end{eqnarray}
This action is conformal invariant if the traceless condition $T_{\mu}^{\mu}=0$ is established. By performing this condition, we should choose $s=(n+1)/4$ to have conformal invariance, where $n$ is dimension of spacetime. Conformal invariance has attracted many people such that 
exact solutions of Einstein-conformal scalar equations are in \cite{Bek1}. Also,
conformally dressed black hole in $(2+1)$ dimensions has been studied in \cite{Mart1}.
Black holes and asymptotics of $(2+1)$ gravity coupled
to a scalar field are investigated in \cite{Henn}. In this paper, black hole solutions that are discussing here, are similar to the black hole ones coupled
to a conformal scalar field in an asymptotically flat $(3+1)$ spacetime.
Topological black holes with a conformally coupled scalar field and electric charge have been also probed \cite{Mart2}.\\
Now, we would like to consider a general value for parameter
"s" in conformal invariance power-law Maxwell lagrangian in the presence of third order quasi topological gravity to get interesting results. Quasi-topological gravity is one type of modified gravities which was introduced to solve the defects of Einstein's gravity. Also, in the low energy limit of string theory, some corrections with higher order curvature are added to the action which gets to Einstein's equations. So, as Einstein's gravity is not proper for strong gravities, this caused to generalize general relativity theory by modified gravities like Love-Lock \cite{Friedman,Schleich,Jacobs} and quasi-topological \cite{Cai2,Mann1,Lemos1,Lemos2,Lemos3,Brenna,Dehghani1,Dehghani2,Dehghani3,Aminn} gravities. These two gravities are almost the same in spherically symmetry conditions, but quasi-topological gravity has the ability to produce effective gravitational effects in less dimensions than Lovelock gravity. For example, when the cubic Lovelock gravity is established in seven and higher dimensions, the quasi-topological gravity is effective in five and higher dimensions.\\
Some attractive subjects have been studied about quasi topological gravity. For example, Black holes in quasi-topological gravity are studied \cite{Myer1}. Quasi-topological gravity in the spherically symmetric case is in \cite{Bazr}. There is also a review of quartic quasi-topological black holes in the presence of the nonlinear electromagnetic Born-Infeld field in \cite{Ghanaa2}. Also, quasi-topological gravity has been investigated in Lifshitz spacetime for general value $z$ in \cite{Ghanaa1} and for $z=1$ in \cite{abkar}. Now, we are eager to construct lifshitz black hole solutions with the third order quasi-topological gravity in the presence of a power-law Maxwell field. \\
In what follows, we first consider the five dimensional action with the third order quasi-topological gravity in the presence of conformal invariance power-law Maxwell field with general parameter "s" in section \ref{Field}. Then, we vary the action in spherically symmetry Lifshitz spacetime with $z=1$ and find the solutions. In section \ref{physical}, we investigate the obtained solutions in asymptotically de sitter, anti-de sitter and flat spacetime. Then, we obtain thermodynamic quantities of this black hole and check the first law of thermodynamics in \ref{thermo}. To have a finite value for total mass of the black hole, we should apply a condition on parameter $s$. We also probe thermal stability of this black hole in \ref{stability}. At last and in section \ref{result}, we have a review on what important notes we have obtained in the whole paper as results. 
\section{main structure of the black hole and its solutions}\label{Field}
Our theory begins with a five dimensional action in the third order quasi-topological gravity that is coupled to the nonlinear power-law Maxwell field with a general power $s$
\begin{equation}\label{Act1}
I_{G}=\int{d^{5}x\sqrt{-g}\bigg\{-2\Lambda+{\mathcal L}_1+\frac{\lambda L^2}{2}{\mathcal L}_2+\frac{7\mu L^4}{4}{\mathcal L}_3+\bigg(-\frac{1}{4}F_{\mu\nu}F^{\mu\nu}\bigg)^{s}\bigg\}},
\end{equation}
where $\Lambda$ is the cosmological constant. ${\mathcal L}_1$, ${\mathcal L}_2$ and ${\mathcal L}_3$ are respectively, Einstein-Hilbert, second order Lovelock (Gauss-Bonnet) Lagrangian and third order corrections in quasi-topological gravity that are defined as
\begin{eqnarray}
{{\mathcal L}_1}=R,
\end{eqnarray}
\begin{eqnarray}
{{\mathcal L}_2}=R_{abcd}R^{abcd}-4R_{ab}R^{ab}+R^2,
\end{eqnarray}
 \begin{eqnarray}
{{\mathcal L}_3}&=&
R_a{{}^c{{}_b{{}^d}}}R_c{{}^e{{}_d{{}^f}}}R_e{{}^a{{}_f{{}^b}}}+\frac{1}{(2n-1)(n-3)} \bigg[\frac{3(3n-5)}{8}R_{abcd}R^{abcd}R-3(n-1)R_{abcd}R^{abc}{{}_e}R^{de}\nonumber\\
&&+3(n+1)R_{abcd}R^{ac}R^{bd}+6(n-1)R_a{{}^b}R_b{{}^c}R_{c}{{}^a}-\frac{3(3n-1)}{2}R_a{{}^b}R_b{{}^a}R +\frac{3(n+1)}{8}R^3\bigg].
\end{eqnarray}
In this action, $\lambda$ and $\mu$ are respectively the coefficients of Gauss-Bonnet and third order quasi topological gravities. The last term in this action is also the power-law Maxwell lagrangian which reduces to linear Maxwell lagrangian if we consider $s=1$\cite{BMRN}.\\ 
The electromagnetic field tensor $F_{\mu\nu}$ is defined as $F_{\mu\nu}=\partial_{\mu}A_{\nu}-\partial_{\nu}A_{\mu}$, where $A_{\mu}$ is vector potential and we choose it as 
\begin{eqnarray}\label{h1}
A_{\mu}=h(r)\delta_{\mu}^{0},
\end{eqnarray}
to have static solutions. Our purpose is to seek asymptotically lifshitz solutions in five dimensional spherical symmetry spacetime with $z=1$
\begin{eqnarray}\label{metr}
ds^2=-\frac{r^2}{L^2}f(r)dt^2+\frac{L^2}{r^2 g(r)}dr^2+\frac{r^2}{L^2} (x^2+y^2+z^2),
\end{eqnarray}
where $L$ is a scale factor relating to the cosmological constant and $f(r)$ and $g(r)$ are the metric functions that should be found. This metric has a three dimensional hypersurface with constant curvature zero in volume $V_{3}$.\\ 
Now, we can obtain equations. For this purpose, we first substitute the above data in the action (\ref{Act1}) and then integrate by parts. By varying this action with respect to functions $g(r)$, $f(r)$ and $h(r)$, we get to the equations
\begin{eqnarray}
\Lambda L^2 r^6+6r^6g-6\lambda r^6 g^2-6\mu r^6g^3+g(ln f)^{'}\bigg(\frac{3}{2}r^7-3\lambda r^7 g-\frac{9}{2}\mu r^7 g^2\bigg)=-(2s-1)2^(s-1)r^6 L^2\bigg(\frac{g}{f}\bigg)^{s}(h^{'})^{2s},
\end{eqnarray}

\begin{eqnarray}
12\mu r^3g^3-2\Lambda L^2 r^3-12r^3g+12\lambda r^3 g^2+9\mu r^4g^2g^{'}+6\lambda r^4 g g^{'}-3r^4g^{'}=(2s-1)2^{s}r^3 L^2\bigg(\frac{g}{f}\bigg)^{s}(h^{'})^{2s},
\end{eqnarray}
\begin{eqnarray}
2(1-2s)r h^{"}-rh^{'}(1-2s)[(ln f)^{'}-(ln g)^{'}]-6h^{'}=0.
\end{eqnarray}
By simplifying the above equations, we get to 
\begin{equation}\label{equ1}
(-1+2\lambda g+3\mu g^2)N^{'}=0,
\end{equation}
\begin{equation}\label{equ2}
\bigg\{3r^4\bigg(-\frac{\Lambda}{6}L^2-g+\lambda g^2+\mu g^3\bigg)\bigg\}^{'}=2^s(2s-1)r^3L^2\bigg(\frac{h^{'}}{N}\bigg)^{2s},
\end{equation}
\begin{equation}\label{equ3}
\bigg(r^3\bigg(\frac{h^{'}}{N}\bigg)^{2s-1}\bigg)^{'}=0,
\end{equation}
where $f(r)=N^2(r)g(r)$.\\
Now, we want to earn the solutions of the above equations. First, we go to equation (\ref{equ1}), which shows that N(r) should be a constant and so we consider it $N(r)=1$. Substituting $N(r)=1$ in the third equation and solving it, leads to answer
\begin{equation}\label{At1}
A_{t}=\left\{
\begin{array}{ll}
$$\frac{(2s-1)}{(2s-4)}qr^{\frac{(2s-4)}{(2s-1)}}$$,\quad \quad\quad\quad \quad\quad\quad  \ {\frac{1}{2}< s<2,}\quad &  \\ \\
$$q \rm ln(r)$$,\quad\quad\quad\quad\quad\quad\quad\quad\quad\quad\quad  \ {s=2,}\quad & 
\end{array}
\right.
\end{equation}
where $q$ is a constant of integration relating to the electric charge of the black hole.
It is obvious that by inserting $s=1$ in $A_{t}$, we get to the vector potential of third order quasi-topological black hole in the presence of Maxwell field \cite{BMRN}.
If we use $N=1$ and equation (\ref{At1}) in equation (\ref{equ2}), it leads to
\begin{eqnarray}\label{equasli}
\mu g^3+\lambda g^2-g+\kappa=0,
\end{eqnarray}
where $\kappa$ is
\begin{equation}\label{h2}
\kappa=\mu_{0}-\frac{M}{3r^4}-L^2(2q^2)^{s}r^{\frac{6s}{1-2s}}\left\{
\begin{array}{ll}
$$\frac{(2s-1)^2}{3(2s-4)}$$,\quad \quad\quad\quad \quad\quad\quad  \ {\frac{1}{2}< s<2,}\quad &  \\ \\
$$ \rm ln(r)$$,\quad\quad\quad\quad\quad \quad\quad\quad\quad  \ {s=2,}\quad & 
\end{array}
\right.
\end{equation}
and $\mu_{0}=-\frac{\Lambda}{6}L^2$ and $M$ is a constant of integration relating to the mass of the black hole. By solving Eq. (\ref{equasli}), we lead to answer
\begin{eqnarray}
g(r)=-\frac{\lambda}{3\mu}-\frac{1}{12\mu}\bigg[\bigg(j(r)+\sqrt{\Gamma+J^2(r)}\bigg)^{\frac{1}{s}}-\frac{\Gamma^{\frac{1}{s}}}{\bigg(j(r)+\sqrt{\Gamma+J^2(r)}\bigg)^{\frac{1}{s}}}\bigg],
\end{eqnarray}
where $\Gamma=-(16(3\mu+\lambda^2))^3$ and
\begin{eqnarray}
J(r)=36[\frac{16}{9}\lambda^3+8\mu\lambda-4\Lambda\mu^2L^2-8\mu^2\frac{M}{r^4}-4\mu^2\frac{(2s-1)^2}{s-2}L^2(2q^2)^sr^{\frac{6s}{1-2s}}],
\end{eqnarray}
which for $s=2$, the last term of $J(r)$ is $-96\mu^2L^2q^4\frac{ln(r)}{r^4}$.\\
The function $g(r)$ is real if and only if $\mu<-\frac{\lambda^2}{3}$.
It should be noted that we have used the condition $\frac{1}{2}< s \leq 2$ for the solutions, which we will explain about it in the last section \ref{thermo}.

\section{behavior of the solutions } \label{physical}
Now, we can investigate the behavior of this third order quasi-topological black hole coupled to the power-law Maxwell field by studying the physical structure of its solutions. For this aim, we first study the electrical field $E(r)$ of this black hole and then we study properties of function $g(r)$ in an asymptotically anti-de sitter(AdS), de sitter(dS) and flat spacetime. we choose $L=1$ for simplicity.
\subsection{electrical field $E(r)$}
\begin{figure}
\center
\includegraphics[scale=0.5]{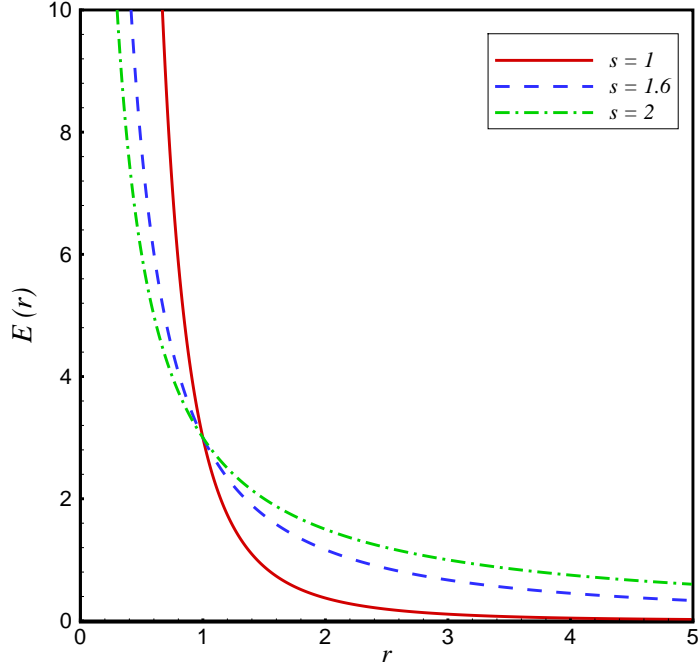}
\caption{\small{Electrical field $E(r)$ versus $r$ for $s=1$(solid red), $s=1.6$(dashed blue) and $s=2$(dashdot green).} \label{Fig1}}
\end{figure}

The electrical field $E(r)$ is obtained as 
\begin{eqnarray}
E(r)=F_{tr}=qr^{\frac{-3}{2s-1}},
\end{eqnarray}  
where for $s=1$, it is reduced to the electrical field of Maxwell theory in third order quasi-topological gravity \cite{BMRN}.
As we said, one motivation for introducing nonlinear electrodynamics was to reduce the divergence of the electrical field of linear Maxwell theory at the origin. To show this benefit, we have plotted $E(r)$ versus $r$ in Fig. (\ref{Fig1}).
In this figure, the electrical field of Maxwell theory(s=1) is in solid red diagram and the electrical fields of power-law Maxwell theory are in blue dash ($s=1.6$) and green dashdot diagrams ($s=2$). This figure shows that for three diagrams, $E(r)$ goes to infinity as $r\rightarrow 0$, but it is clear that in power-law Maxwell cases, the diagrams have a less divergence in $r=0$ than the one in Maxwell case. So, power-law Maxwell nonlinear lagrangian has the ability to reduce the divergence of $E(r)$ at the origin. Also, by increasing the value of parameter $s$, this divergence becomes weaker.
For $r\rightarrow \infty$, the electrical field of this black hole goes to 0 independent to the value of parameter $s$.\\
\subsection{function $f(r)$}
\begin{figure}
\center
\includegraphics[scale=0.5]{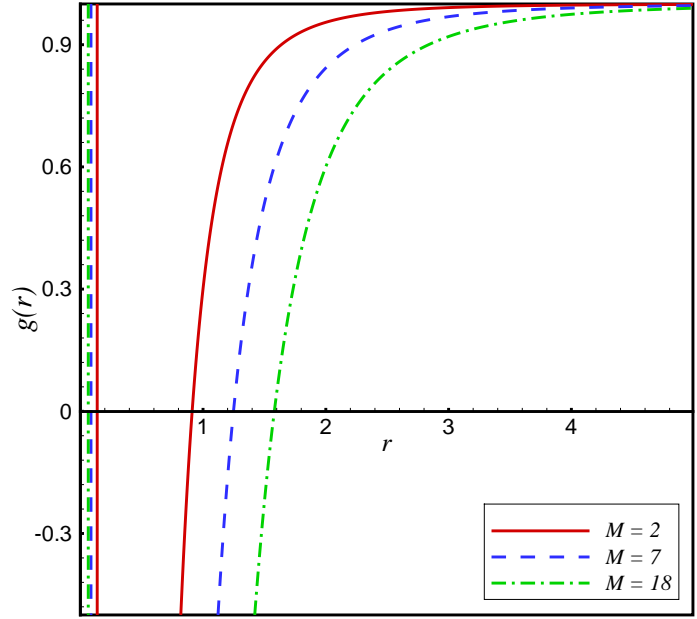}
\caption{\small{Asymptotically AdS solution $f(r)$ versus $r$ for different $M$ with $s=0.9$, $q=0.1$, $\lambda=0.04$ and  $\mu=-0.001$.} \label{Fig2}}
\end{figure}

\begin{figure}
\center
\includegraphics[scale=0.5]{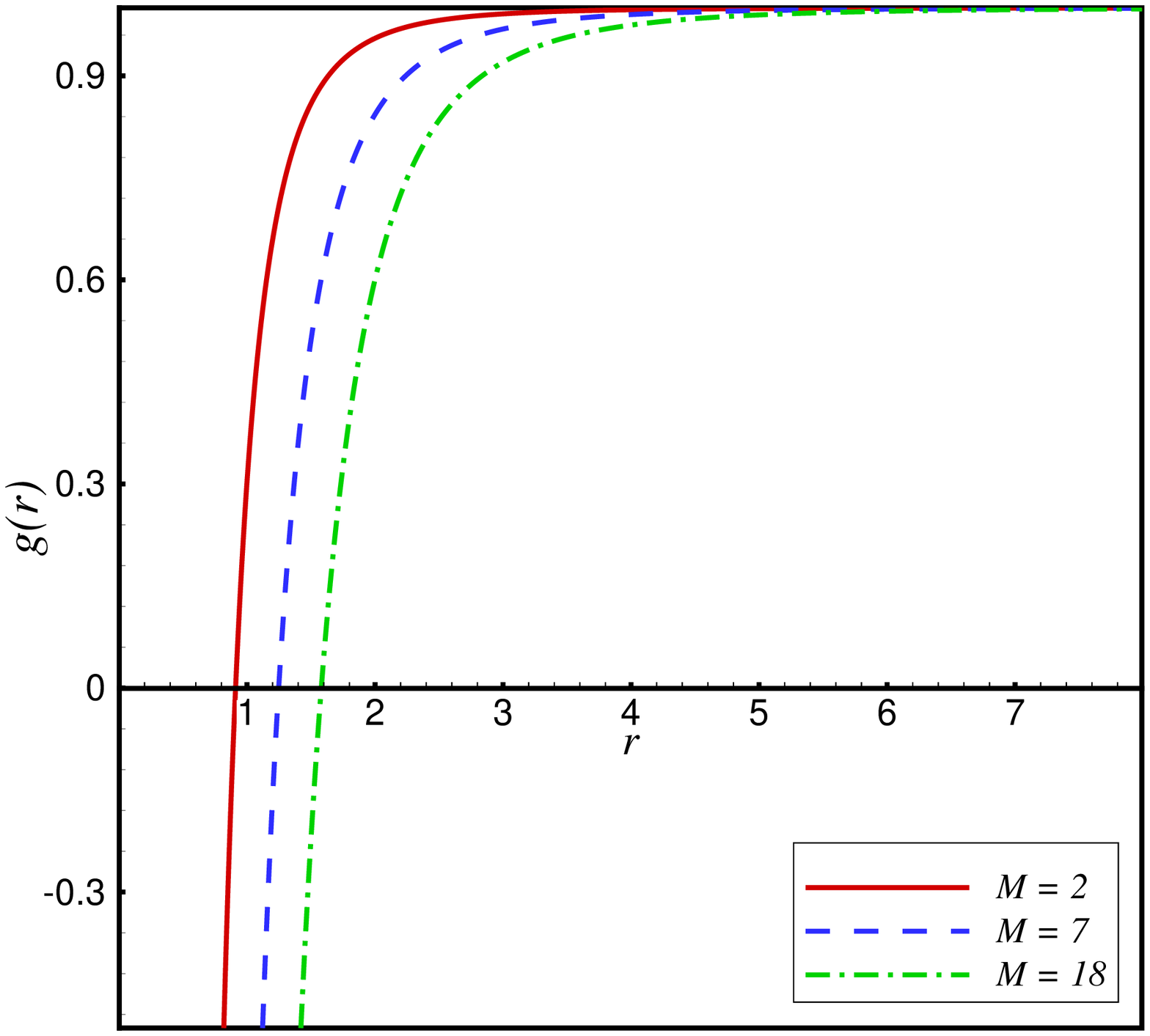}
\caption{\small{Asymptotically AdS solution $f(r)$ versus $r$ for different $M$ with $s=1.8$, $q=0.1$, $\lambda=0.04$ and $\mu=-0.001$.} \label{Fig3}}
\end{figure}

\begin{figure}
\center
\includegraphics[scale=0.5]{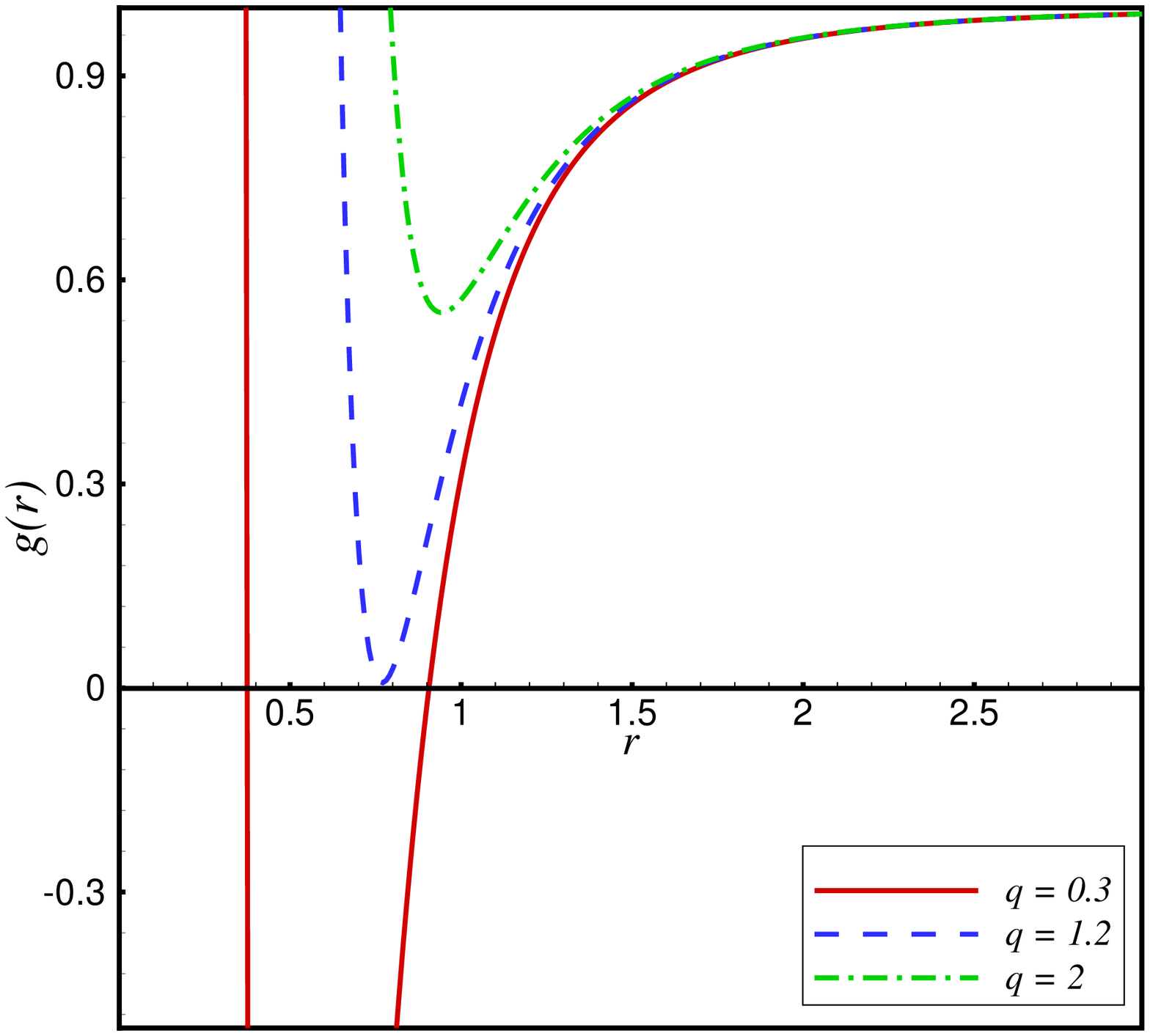}
\caption{\small{Asymptotically AdS solution $f(r)$ versus $r$ for different $q$ with $s=0.8$, $M=2$, $\lambda=0.04$ and $\mu=-0.001$.} \label{Fig4}}
\end{figure}

\begin{figure}
\center
\includegraphics[scale=0.5]{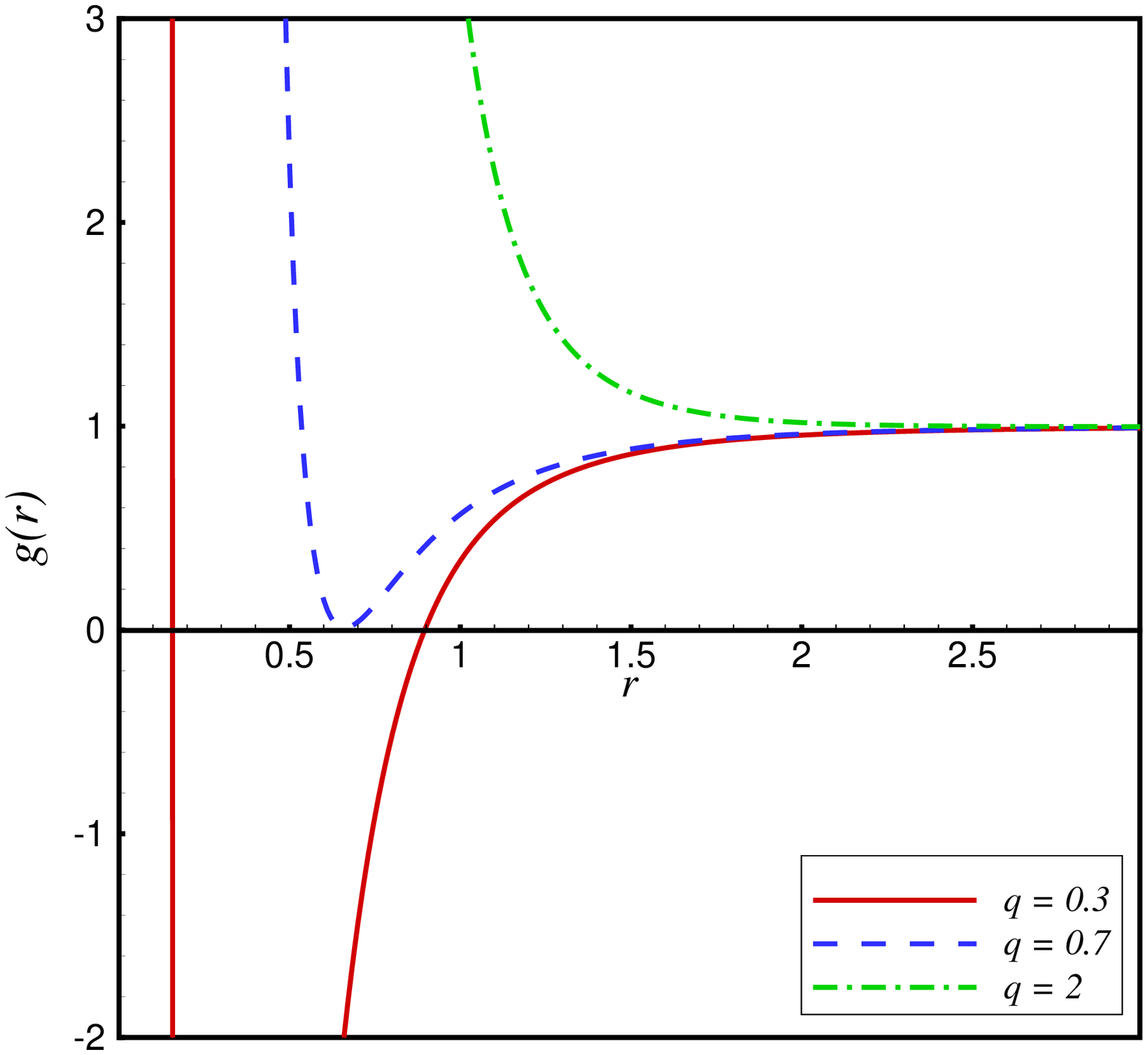}
\caption{\small{Asymptotically AdS solution $f(r)$ versus $r$ for different $q$ with $s=1.1$, $M=2$, $\lambda=0.04$ and $\mu=-0.001$.} \label{Fig5}}
\end{figure}

\begin{figure}
\center
\includegraphics[scale=0.5]{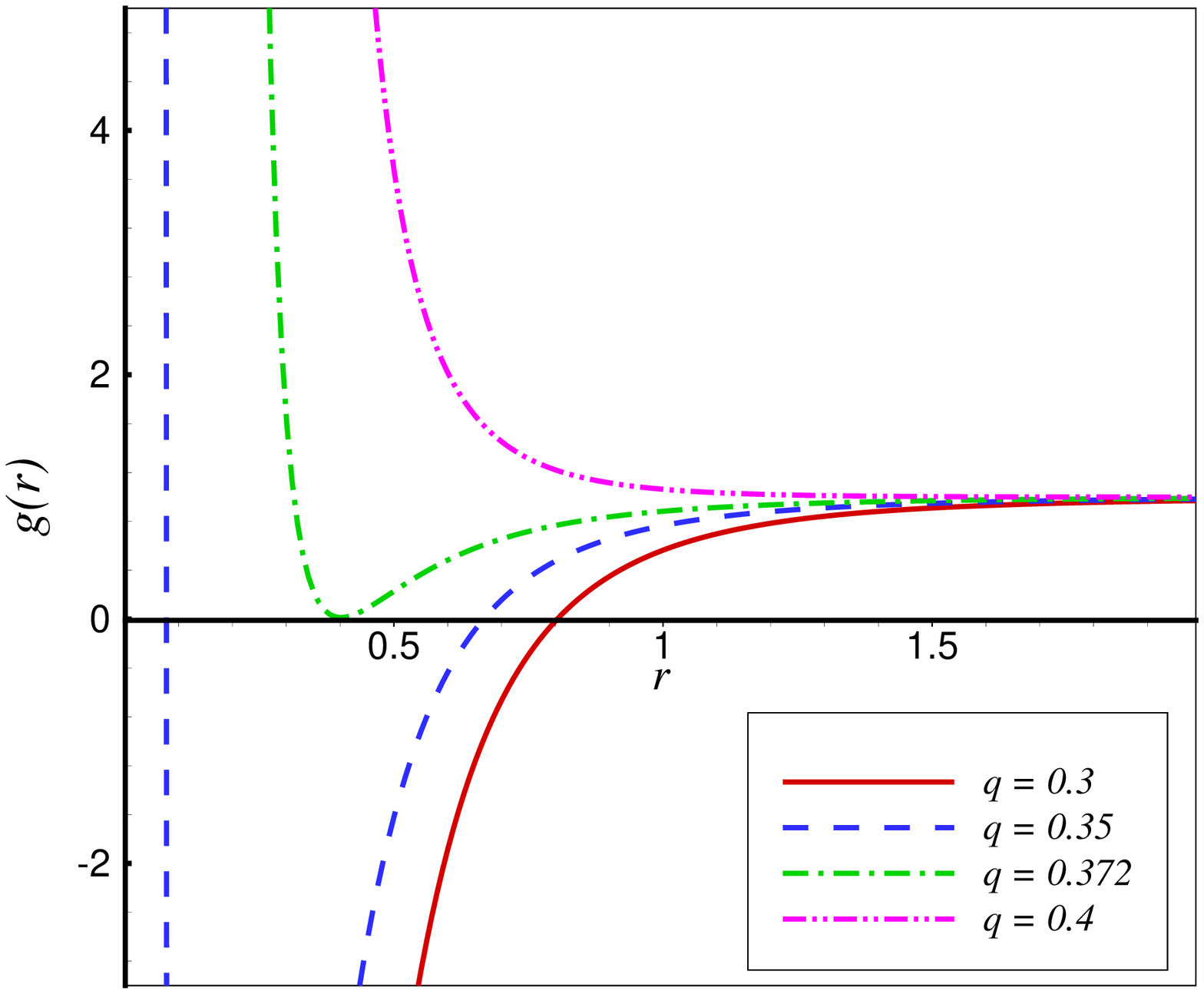}
\caption{\small{Asymptotically AdS solution $f(r)$ versus $r$ for different $q$ with $s=1.8$, $M=2$, $\lambda=0.04$ and $\mu=-0.001$.} \label{Fig6}}
\end{figure}

\begin{figure}
\center
\includegraphics[scale=0.5]{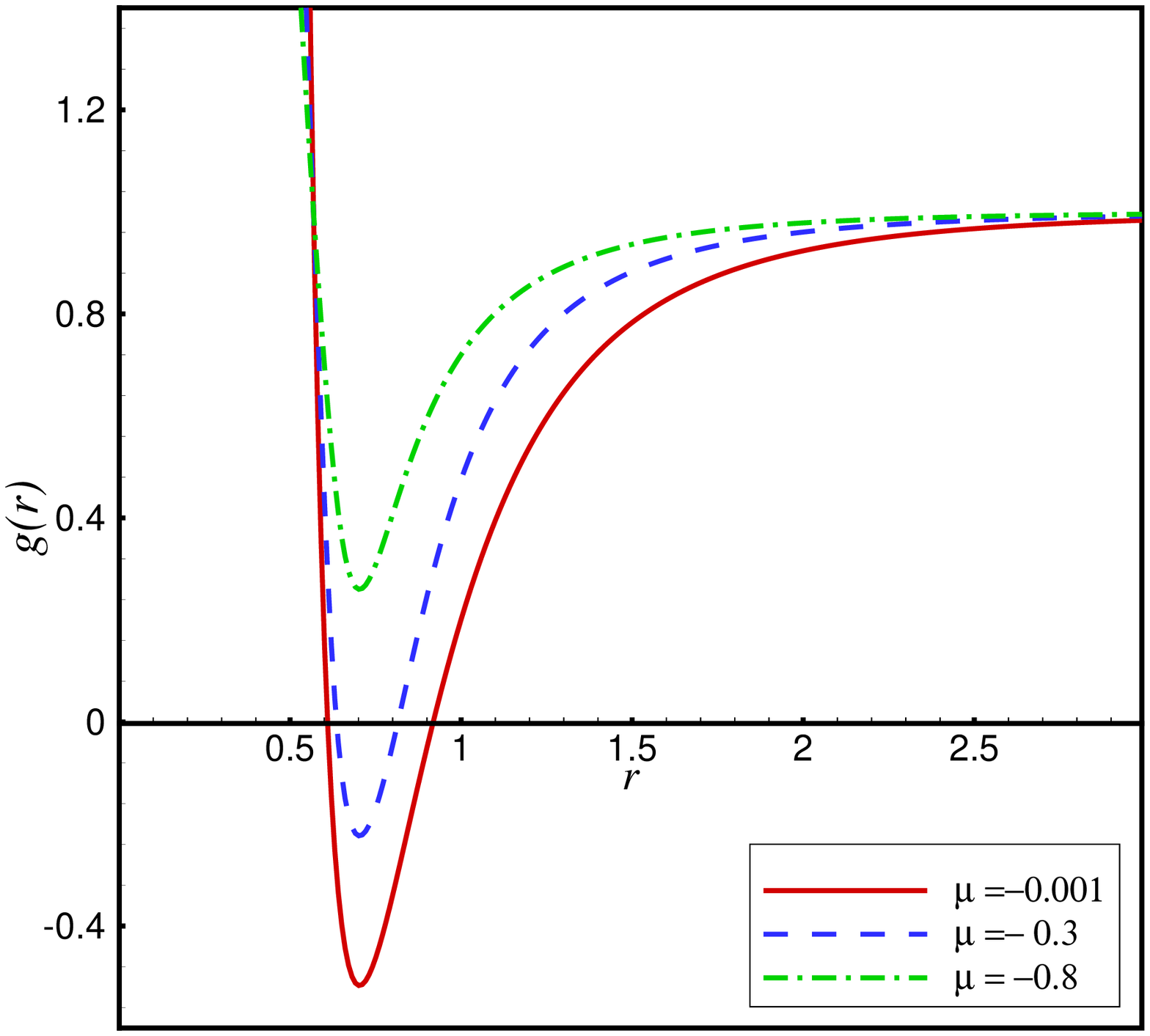}
\caption{\small{Asymptotically AdS solution $f(r)$ versus $r$ for different $\mu$ with $s=1.1$, $q=1$, $M=4$ and $\lambda=0.04$.} \label{Fig7}}
\end{figure}

Now, we want to explore behavior of $g(r)$ versus $r$ in Ads, dS and flat spacetime. The difference of these spacetimes is in value $\Lambda$ which we will study. It is notable that the function $g(r)$ is real for the values $\lambda=0.04$ and $\mu=-0.001$, because they satisfy the condition $\mu<-\frac{\lambda^2}{3}$. So we use these two parameters for three Ads, dS and flat apacetimes. \\
We first start with AdS solutions. Applying condition
\begin{eqnarray}\label{conditionAds1}
lim_{r\rightarrow\infty}g(r)=1
\end{eqnarray}
in Eq.(\ref{equasli}), results to bellow cosmological constant 
\begin{eqnarray}\label{conditionAds}
\Lambda=-\frac{6(1-\mu-\lambda)}{L^2}.
\end{eqnarray}
Cosmological constant is negative in AdS spacetime($\Lambda<0$). So by exerting $\Lambda<0$ in Eq.(\ref{conditionAds}), we get to $\mu+\lambda+c<1$, which is necessary for the solutions of third order quasi-topological black hole coupled to the power-law Maxwell theory in AdS spacetime.\\
We have plotted figures (\ref{Fig2})-(\ref{Fig7}) to study the behavior of the solutions in AdS spacetime. As we can see in all these figures, the condition (\ref{conditionAds1}) is confirmed and $g(r)$ goes to $1$ , as $r$ goes to infinity. In Fig. (\ref{Fig2}), we have investigated $g(r)$ versus $r$ for different values of $M$ and constant values of $q$ and $s$. Independent to the value of $M$ and for fixed parameters $s=0.9$ and $q=0.1$, the function $g(r)$ goes to positive infinity as $r\rightarrow 0$. For these parameters, the figure shows a black hole with inner and outer horizons for all value of $M$. Horizons are the positive roots of the equation $g(r)=0$. $r_{+}$ is defined as the radial coordinate of the outermost horizon of the black hole which is positive root of the equation $g(r_{+})=0$. The value of the inner horizon is independent to the value of M, while the value of the outer one is related to $M$ and grows by increasing the value of $M$. \\
In fig. (\ref{Fig3}), we have repeated the parameters in fig. (\ref{Fig2}) but with a different value for $s$. We have considered $s=1.8$ for this figure. It is clear that by changing parameter $s$, the behavior of $g(r)$ changes and there is just one horizon for this black hole which its value increases as the value of parameter $M$ increases. Also, with comparison to fig.(\ref{Fig2}), $g(r)$ goes to negative infinity for small value $r$, .\\
In fig.(\ref{Fig4}), we have plotted $g(r)$ versus $r$ for different values of $q$ and fixed parameters $s=0.8$ and $M=2$. In this figure, $g(r)\rightarrow +\infty$ as $r\rightarrow 0$. Depending on the value of $q$, we can have a black hole with two horizons (for $q=0.3$), an extreme black hole (for $q=1.2$) and a naked singularity (for $q=2$). So, we can derive that for these values of parameters $s$ and $M$, there is a $q_{\rm ext}$ that the theory defines an extreme black hole for $q=q_{\rm ext}$, a black hole with two horizons for $q<q_{\rm ext}$ and a naked singularity for $q>q_{\rm ext}$. Therefore, for these parameters, small $q$ leads to a black hole with two horizons and by increasing the value of $q$, the positive region of $g(r)$ becomes more as for $q=2$, $g(r)$ is positive for all values of $r$.\\
In fig.(\ref{Fig5}), we have acted like in fig. (\ref{Fig4}) and for different values of $q$ but for fixed parameters $s=1.1$ and $M=2$. It is obvious that $g(r)$ has a similar behavior in both figures but there is a difference between them. As the parameter $s$ increases relative to the previous figure, the value of $q_{\rm ext}$ becomes smaller.\\
In fig. (\ref{Fig6}), we have plotted $g(r)$ versus $r$ for different values of $q$ and the same parameters as in figures (\ref{Fig4}) and (\ref{Fig5}) but for different value $s=1.8$.
As we see in this figure, there is a $q_{\rm ext_{2}}$ that we have an extreme black hole for $q=q_{\rm ext_{2}}$ and a naked singularity for $q>q_{\rm ext_{2}}$. Also for $q<q_{\rm ext_{2}}$, there is a $q_{min}$ which $_{min}<q<q_{\rm ext_{2}}$ is related to a black hole with inner and outer horizons and $q<q_{min}$ describes a nonextreme black hole. Summarizing the contents of figures (\ref{Fig4}), (\ref{Fig5}) and (\ref{Fig6}) demonstrates that for constant values of $M$, $\lambda$ and $\mu$, there is a $s_{max}$ which for $s<s_{max}$, there is a black hole with two horizons for $q<q_{ext}$, a naked singularity for $q>q_{ext}$ and an extreme black hole for $q=q_{ext}$. Also in this region of $\frac{1}{2}<s<s_{max}$, as the parameter $s$ increases, the value of $q_{ext}$ decreases. For $s>s_{max}$, there is a $q_{ext_{2}}$ that $q=q_{ext_{2}}$ and $q>q_{ext_{2}}$ respectively describe an extreme black hole and a naked singularity. Also, for $q<q_{ext_{2}}$, there is a $q_{min}$ which $q_{min}<q<q_{ext_{2}}$ shows a black hole with two horizons and $q<q_{min}$ characterizes a nonextreme black hole. We should also say that the value of $q_{ext_{2}}$ is smaller then the value of $q_{ext}$.\\
In fig.(\ref{Fig7}), we have studied the behavior of $g(r)$ for constant parameters $s=1.1$, $q=1$, $M=4$, $\lambda=0.04$ and for different values of parameter $\mu$. We can deduce from this figure that there is a $\mu_{max}$ that $\mu<\mu_{max}$ leads to a naked singularity and $\mu>\mu_{max}$ results to a black hole with two horizons. So for the above fixed parameters, we can have a black hole with two horizons if $\mu<0$ and its value $|\mu|$ is small near $0$.       
  

\begin{figure}
\center
\includegraphics[scale=0.5]{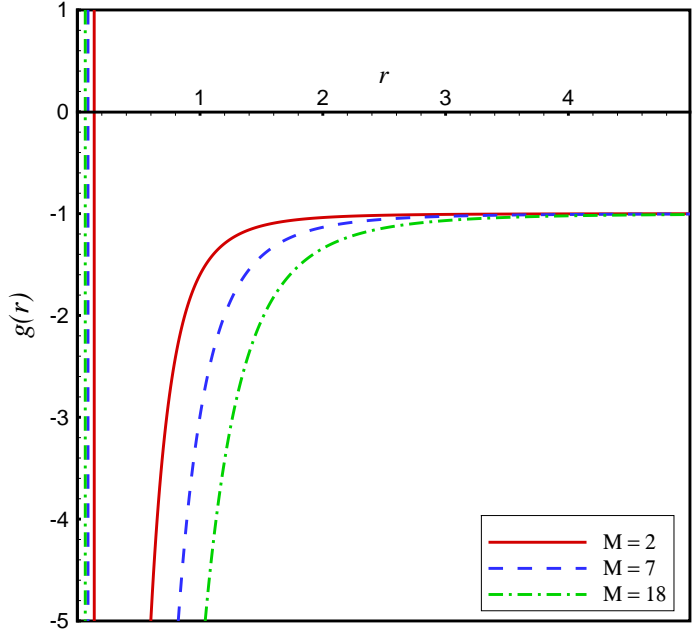}
\caption{\small{Asymptotically dS solution $f(r)$ versus $r$ for different $M$ with $s=0.9$, $q=0.1$, $\lambda=0.04$ and $\mu=-0.001$.} \label{Fig8}}
\end{figure}

\begin{figure}
\center
\includegraphics[scale=0.5]{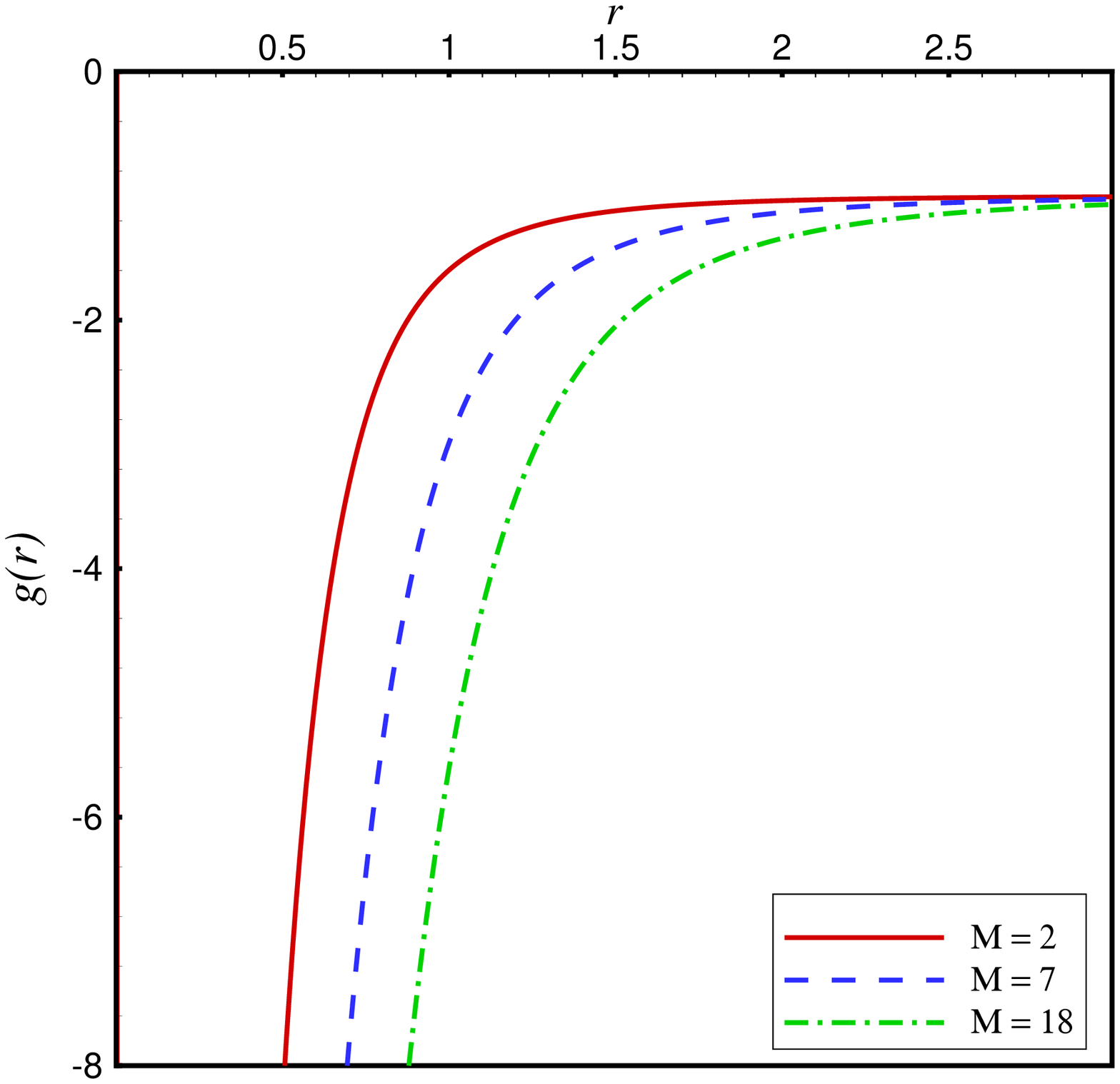}
\caption{\small{Asymptotically dS solution $f(r)$ versus $r$ for different $M$ with $s=1.8$, $q=0.1$, $\lambda=0.04$ and $\mu=-0.001$.} \label{Fig9}}
\end{figure}

\begin{figure}
\center
\includegraphics[scale=0.5]{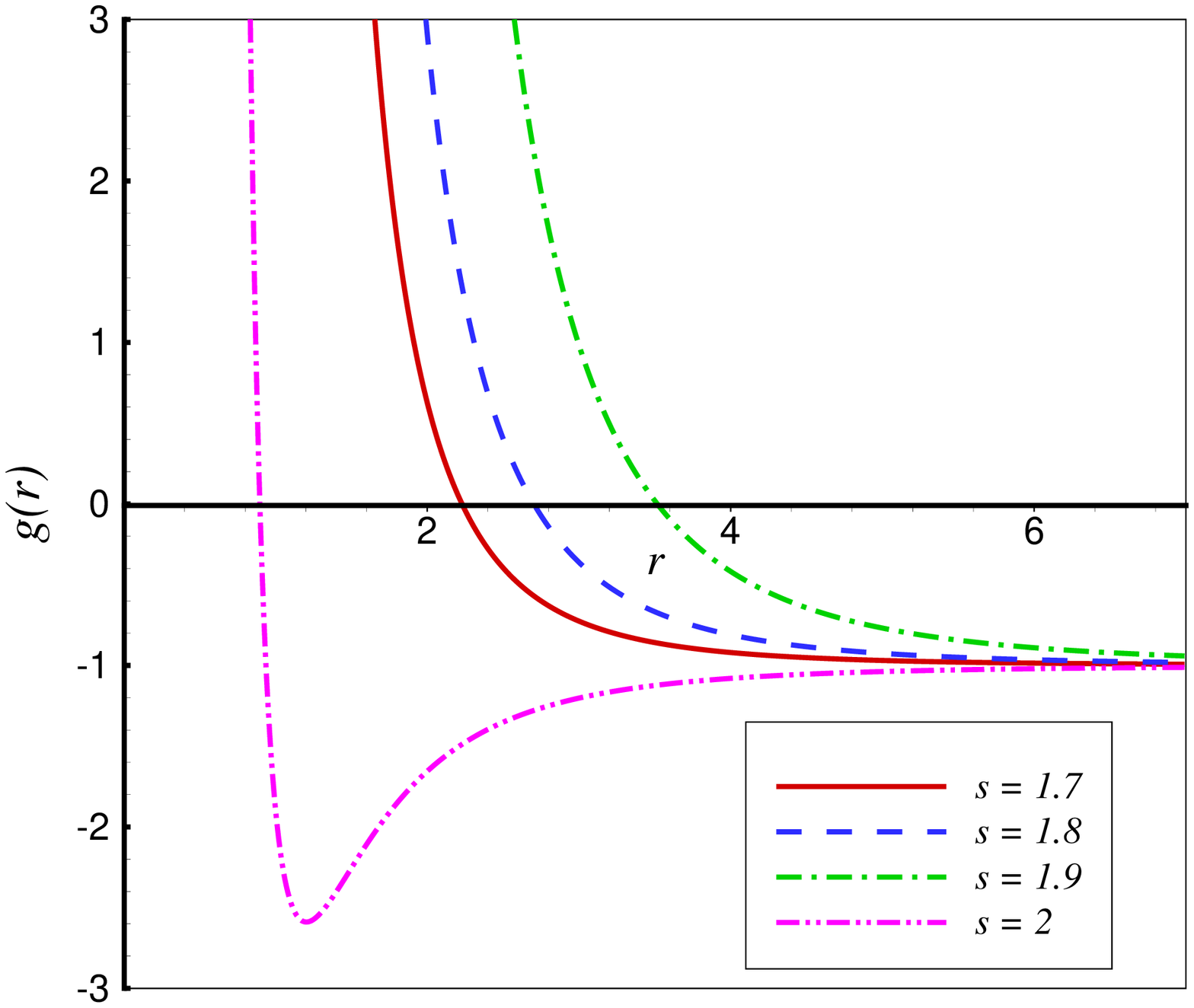}
\caption{\small{Asymptotically dS solution $f(r)$ versus $r$ for different $s$ with $q=1.4$, $M=3$, $\lambda=0.04$ and $\mu=-0.001$.} \label{Fig10}}
\end{figure}

\begin{figure}
\center
\includegraphics[scale=0.5]{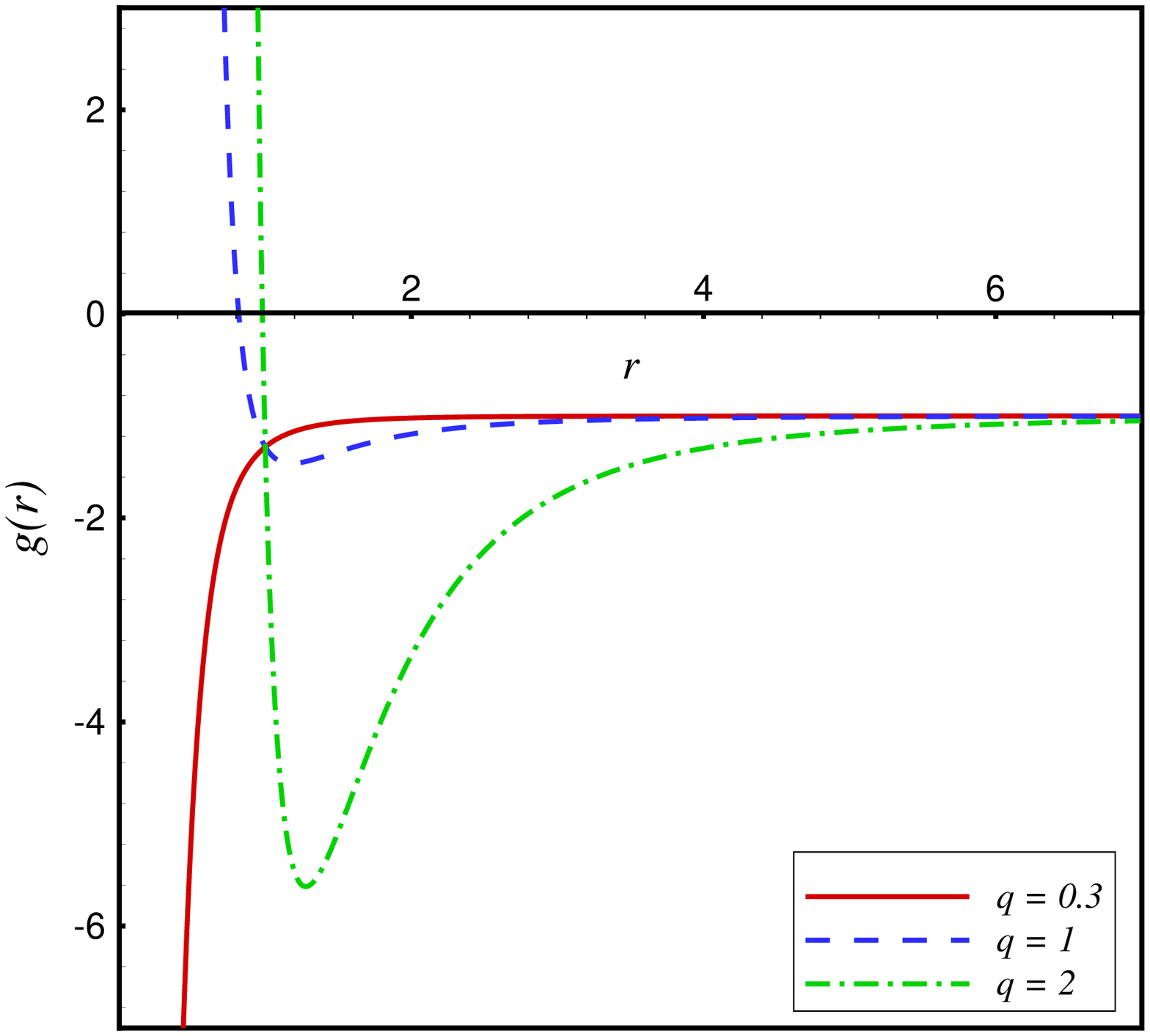}
\caption{\small{Asymptotically dS solution $f(r)$ versus $r$ for different $q$ with $s=2$, $M=1$, $\lambda=0.04$ and $\mu=-0.001$.} \label{Fig11}}
\end{figure}

The different behavior of function $g(r)$ in AdS and dS spacetime caused us to probe its behavior separately. By exerting the condition
\begin{eqnarray}\label{gdS}
\lim_{r\rightarrow\infty} g(r)=-1,
\end{eqnarray}
to the relation (\ref{equasli}), we get to the cosmological constant in dS spacetime for this black hole as
\begin{eqnarray}\label{condds}
\Lambda=\frac{6(1-\mu+\lambda)}{L^2}.
\end{eqnarray}    
$\Lambda>0$ is the characteristic of dS spacetime. By applying this to the above equation (\ref{condds}), we get to the requirement $\lambda-\mu>-1$ for dS solutions. To know the behavior of the solutions in dS spacetime better, we have plotted figures (\ref{Fig8})-(\ref{Fig11}) in dS spacetime for $L=1$.\\  
As it is clear, the condition (\ref{gdS}) is obvious in these figures so that the function $g(r)$ goes to $-1$ for large $r$.\\
In fig.(\ref{Fig8}), we have investigated $g(r)$ versus $r$ for different values of $M$ and fixed values of $s=0.9$ and $q=0.1$. It is clear that for the same parameters, general behavior of $g(r)$ in dS spacetime is similar to the ones in dS spacetime. For example, it goes to $+\infty$ for $r\rightarrow 0$, but their difference is that $g(r)$ cuts off axis $r$ just in one point in dS spacetime. So for the same above conditions, dS solutions have just one horizon while there are two horizons in AdS spacetime. \\
Like fig.(\ref{Fig8}), we have probed the behavior of the function for different $M$ and constant $s=1.8$ and $q=0.1$ in fig.(\ref{Fig9}). For the same conditions, the general behavior of this figure is like the one in AdS apacetime, but with a difference. In these conditions, the solutions in dS spacetime don't have any horizons. So, for the same conditions and $q=0.1$, the number of the horizons in ds spacetime is one less than the one in AdS spacetime.   
So, because the general behaviors of AdS solutions are similar to the dS ones, we have avoided to bring the similar behavior of $g(r)$. \\
It seems that $g(r)$ has a different behavior for $s=2$ than the other values of $s$. So, this made us eager to study the behavior of $g(r)$ for constant $M=3$ and $q=1.4$ and for different values of $s=1.7,1.8,1.9,2$ in figure (\ref{Fig10}). As it is seen, $g(r)$ has a regular behavior for $s<2$, so that by increasing the value of $s$, the value of horizons becomes bigger. However, we don't have this regularity for $s=2$. Whereas we expect that the value of the horizon for $s=2$ will be bigger than the ones for other $s$, its value is smaller than all of them.\\
In fig.(\ref{Fig11}), we have also checked out the behavior of $g(r)$ for $s=2$, $M=1$ and different values of $q$ with $s=2$. As we can see, there is a $q_{1}$ that for $q<q_{1}$, $g(r)$ is negative for all values of $r$ and goes to $-\infty$ for $r\rightarrow 0$. But for $q>q_{1}$, our solution has a horizon in $r=r_{ext}$ which $g(r)$ is negative for $r>r_{ext}$ and positive for $r<r_{ext}$. Also, for $q>q_{1}$, $g(r)$ goes to $+\infty$ for $r\rightarrow 0$. It is also worth mentioning that for $q>q_{1}$, by increasing the value of $q$, $r_{ext}$ increases. Also for $r>r_{ext}$, whatever the value of $q$ is bigger, $|g(r)|$ becomes bigger.\\   
Now, we deal with flat solutions in which $\lim_{r\rightarrow\infty} g(r)=0,$. These solutions are obtained by setting $\Lambda=0$. To know more, we have plotted $g(r)$ versus $r$ for constant parameters $\lambda=0.04$, $\mu=-0.001$, $M=3$, $s=1.4$ and different values of $q$. In all diagrams, $g(r)$ goes to $0$ as $r\rightarrow \infty$, while for $r\rightarrow 0$, the behavior of $g(r)$ depends on $q$. In this limit, $g(r)\rightarrow -\infty$ for $q=0.1$ and $g(r)\rightarrow +\infty$ for $q=0.4, 0.7, 1.2$. Also, for two values $q=0.1$ and $q=1.2$, $g(r)$ is respectively negative and positive for all values of $r$ and doesn't cut off the axis $r$. So, in these two cases, we have a naked singularity. But for values $q=0.4$ and $0.7$, $g(r)$ cuts off the axis $r$ and so we have a black hole with one horizon.    
\begin{figure}
\center
\includegraphics[scale=0.5]{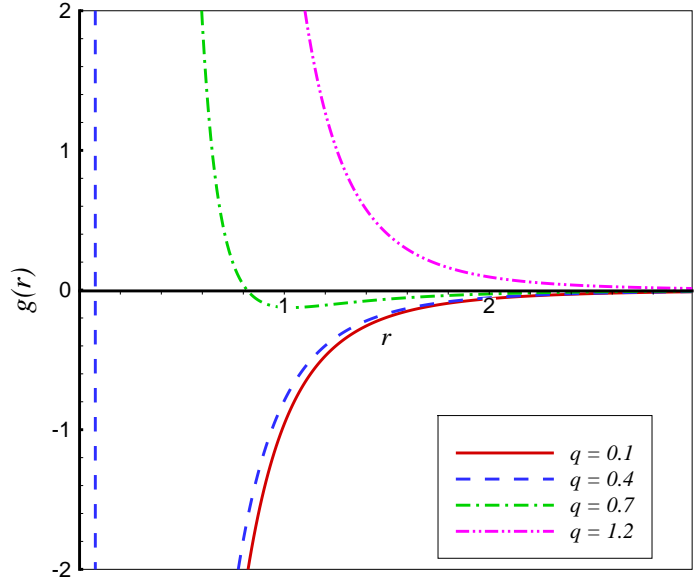}
\caption{\small{Asymptotically flat solution $f(r)$ versus $r$ for different $q$ with $s=1.4$, $M=3$, $\lambda=0.04$ and $\mu=-0.001$.} \label{Fig12}}
\end{figure}
\section{Thermodynamics of the solutions} \label{thermo}
So far, we studied about physical structure of the solutions and now, we are eager to study thermodynamics of the solutions and check the first law of thermodynamics for them. \\
As thermodynamic properties of a black hole are obtained on its event horizon, so the prerequisite for investigating the thermodynamics of a black hole is to search for the event horizon. The event horizon of third order quasi-topological black hole coupled to the power-law Maxwell field is obtained by solving the equation $g(r_{+})=0$ which leads to 
\begin{equation}\label{Mconstant}
M=3\mu_{0}r_{+}^{4}-3L^2(2q^2)^{s}r_{+}^{\frac{4-2s}{1-2s}}\left\{
\begin{array}{ll}
$$\frac{(2s-1)^2}{(2s-4)}$$,\quad \quad\quad\quad \quad\quad\quad  \ {\frac{1}{2}< s<2,}\quad &  \\ \\
$$\rm ln(r_{+})$$,\quad\quad\quad\quad\quad\quad\quad\quad  \ {s=2.}\quad & 
\end{array}
\right.
\end{equation}
Now, we can go to obtain thermodynamic quantities like entropy, temperature and electric potential and conserved quantities such as mass and charge of the black hole. \\
In general, entropy can be defined by the so-called area law which informs that the entropy of a black hole is proportional to one- quarter of its event horizon area \cite{Beck}. This is usable for each black holes, until their theories are in Einstein's gravity. As we go to the modified gravities, we should add some corrections to this base formula. For the obtained solutions, the new formula for the entropy follow as \cite{Wald}
\begin{eqnarray}
S=-2\pi \oint d^{n-1} x \sqrt{\tilde{g}} Y^{abcd} \hat{\epsilon}_{ab}\hat{\epsilon}_{cd},
\end{eqnarray}     
where the integration is on $(n-1)$-dimensional hypersurface. $Y^{abcd}=\frac{\partial \mathcal L}{\partial R_{abcd}}$, where $\mathcal L$ can be lagrangian of Einstein's, Gauss-Bonnet or third order quasi-topological theory. $\tilde{g}_{\mu\nu}$ is the induced metric on the $(n-1)$-spacelike hypersurface which $\tilde{g}$ is its determinant. According to Eq. (\ref{metr}), this metric shows a $(n-1)$-dimensional hypersurface with constant curvature which leads to a zero value for  $Y^{abcd}\hat{\epsilon}_{ab}\hat{\epsilon}_{cd}$ in Gauss-Bonnet and third order gravity. so the entropy of this black hole is obtained as
\begin{eqnarray}\label{entropy}
S=V_{n-1}\frac{r_{+}^3}{4},
\end{eqnarray} 
where $v_{n-1}$ is the volume of $(n-1)$-dimensional hypersurface.\\
Hawking temperature at the event horizon $r_{+}$ is defined as \cite{zebarjad}
\begin{eqnarray}
T_{+}=\frac{\kappa}{2\pi}=\frac{1}{2\pi}\sqrt{-\frac{1}{2}(\nabla_{\mu}\chi_{\nu})(\nabla^{\mu}\chi^{\nu})},
\end{eqnarray} 
where $\kappa$ is the surface gravity and $\chi=B\, \partial/\partial t$ is the null killing vector of the horizon. So, temperature is obtained as
\begin{eqnarray}\label{Temp}
T_{+}=\bigg(\frac{r^2 g^{'}}{4\pi L^2}\bigg)_{r=r_{+}}=-\frac{\Lambda s}{4\pi(2s-1)}r_{+}+\frac{M(s-2)}{6(2s-1)\pi L^2}\frac{1}{r_{+}^8},
\end{eqnarray}
where $B$ is obtained later.
By the standard definition
\begin{eqnarray}\label{potential1}
U=A_{\nu}\chi^{\nu}\mid_{r\rightarrow\infty}-A_{\nu}\chi^{\nu}\mid_{r=r_{+}},
\end{eqnarray}
and Eq. (\ref{At1}), potential is obtained as 
\begin{equation}\label{UU}
U=\left\{
\begin{array}{ll}
$$B \frac{2s-1}{2s-4}q r_{+}^{\frac{2s-4}{2s-1}}$$,\quad \quad\quad\quad \quad\quad\quad  \ {\frac{1}{2}< s<2,}\quad &  \\ \\
$$Bq ln(r_{+})$$,\quad\quad\quad\quad\quad\quad\quad\quad\quad\quad\quad  \ {s=2.}\quad & 
\end{array}
\right.
\end{equation}
Electric charge per unit volume $V_{n-1}$ is also obtained by using Gauss law
\begin{eqnarray}\label{charge2}
Q=\frac{2^{s-1}}{4\pi} q^{2s-1}.
\end{eqnarray}
Also, by writing metric (\ref{metr}) in the form of the reference background metric, the ADM (Arnowitt-Deser-Misner) energy of the black hole per unit volume $V_{n-1}$ is obtained at the limit $r\rightarrow\infty$ as \cite{abkar}
\begin{eqnarray}\label{mass1}
m=\frac{M}{16\pi L^2}.
\end{eqnarray}
After obtaining the conserved and thermodynamic quantities, we want to check out the first law of thermodynamics for the solutions. By using the equations (\ref{entropy}), (\ref{charge2}), (\ref{mass1}) and (\ref{Mconstant}), we first obtain m-smarr formula as
\begin{equation}\label{msmar}
m(S, Q)=-\frac{\Lambda}{32\pi}(4S)^{\frac{4}{3}}-\frac{2^{s}}{16\pi}\bigg(\frac{4\pi Q}{2^{s-1}}\bigg)^{\frac{2s}{2s-1}}\bigg(4S\bigg)^{\frac{4-2s}{3(1-2s)}}\left\{
\begin{array}{ll}
$$\frac{(2s-1)^2}{(2s-4)}$$,\quad \quad\quad\quad \quad\quad\quad  \ {\frac{1}{2}< s<2,}\quad &  \\ \\
$$ln(4S)$$,\quad\quad\quad\quad\quad\quad\quad\quad\quad\quad\quad  \ {s=2.}\quad & 
\end{array}
\right.
\end{equation}
Now, if we regard $S$ and $Q$ as a complete set of extensive parameters for the mass $m(S,Q)$, then their intensive
parameters are respectively defined as
\begin{eqnarray}\label{first}
\bigg(\frac{\partial m}{\partial S}\bigg)_{Q}=T\,\,\,\,\,,\bigg(\frac{\partial m}{\partial Q}\bigg)_{S}=U.
\end{eqnarray}
Calculating the above quantities shows that they are exactly the same as the ones in Eqs. (\ref{Temp}) and (\ref{UU}), if and only if we choose $B=s$. So, these results show that the obtained solutions for this black hole satisfy the first law of thermodynamics  
\begin{eqnarray}\label{firstlaw}
dm= TdS+UdQ.
\end{eqnarray}
Eq. (\ref{firstlaw}) shows that the total mass $m$ is linearly related to $U$. On the other hand, $U$ is proportional to $A_{t}$ according to the relation (\ref{potential1}). Therefore, to obtain a finite value for the mass, it is necessary that $A_{t}$ has a finite value at the infinity which leads to the condition $\frac{1}{2}< s \leq 2$. We used this condition in the prior parts.   

\section{Thermal stability}\label{stability}
To have a better viewpoint of this black hole, it is good to study thermal stability of this black hole in grand canonical ensemble. For this purpose, we should consider $S$ and $Q$ as extensive parameters. We can reach to the thermal stability, if for small variations of parameters $S$ and $Q$, the mass of the black hole $m(S,q)$ will be a convex function of its extensive variables. To reach this stability in grand canonical ensemble, the determinant of Hessian matrix $(H)$ should be positive where 
\begin{eqnarray}
H=\left[
\begin{array}{ccc}
\Big(\frac{\partial ^2 m}{\partial S^2}\Big) & \Big(\frac{\partial ^2 m}{\partial S\partial Q}\Big)\\
\Big(\frac{\partial ^2 m}{\partial Q\partial S}\Big) & \Big(\frac{\partial ^2 m}{\partial Q^2}\Big)
\end{array} \right].
\end{eqnarray}
Using the fact that $\Big(\frac{\partial ^2 m}{\partial S\partial Q}\Big)=\Big(\frac{\partial ^2 m}{\partial Q\partial S}\Big)$, the determinant of the Hessian matrix(we call it det(H) for easiness.) is obtained as
\begin{eqnarray}
det(H)=-\frac{s 2^{s} r_{+}^{\frac{2+2s}{1-2s}}\bigg(\frac{4\pi Q}{2^{s-1}}\bigg)^{\frac{2s}{2s-1}}\bigg[2^{s} \bigg(s-\frac{1}{2}\bigg)^2 r_{+}^{\frac{6s}{1-2s}}\bigg(\frac{4\pi Q}{2^{s-1}}\bigg)^{\frac{2s}{2s-1}}-\frac{\Lambda}{2}\bigg]}{36\pi^2(s-2)Q^2}.
\end{eqnarray}
The other tip which should be considered is that negative temperature is not physical. So, both conditions $det(H)>0$ and $T>0$ should be satisfied for establishing thermal stability. By using equations \eqref{Mconstant} and \eqref{charge2}, we have simplified temperature in Eq. \eqref{Temp} to   
\begin{eqnarray}
\Big(\frac{\partial m}{\partial S}\Big)=T=-\frac{r_{+}}{6\pi}\bigg[2^{s} \bigg(s-\frac{1}{2}\bigg) r_{+}^{\frac{6s}{1-2s}}\bigg(\frac{4\pi Q}{2^{s-1}}\bigg)^{\frac{2s}{2s-1}}+\Lambda\bigg].
\end{eqnarray}
If we use the mentioned condition $\frac{1}{2}< s\leq 2$, the above temperature is negative for dS and flat spacetimes (with characteristics $\Lambda>0$ and $\Lambda=0$ respectively). So this black hole doesn't have thermal stability for dS and flat solutions.
As $\Lambda$ is negative for AdS solutions, $det(H)$ is positive for all values of $r_{+}$ independent to the values of the other parameters. So, thermal stability of third order quasi-topological black hole coupled to the power-law Maxwell field is established just for Ads spacetime with the necessity $T>0$.

\section{concluding results}\label{result} 
Linear Maxwell theory has some problems the most of which are important:\\
1- electrical field of this theory diverges in $r=0$,\\
2- it is not conformal invariance in higher dimensions.\\
The first problem caused to introduce nonlinear electrodynamics theory which includes power-law Maxwell source. We showed in this paper that power-law Maxwell theory can reduce the divergence of the electrical field at the origin that is caused by linear Maxwell theory. Power-law Maxwell theory can also establish conformal invariance in higher dimensions with traceless condition $T_{\mu}^{\mu}=0$.\\
This attractive theory made us eager to consider power-law Maxwell theory in third order quasi-topological gravity. So, by considering a five dimensional action in third order quasi-topological gravity with power-law Maxwell field and varying it, we could get to gravitational and material equations. We then solved the equations and obtained solutions. The obtained solutions are finite at infinity if $\frac{1}{2}<s\leq 2$ and they are real if $\mu<-\frac{\lambda^2}{3}$. \\
Then, we investigated behavior of electrical field $E(r)$ and physical structure of the solutions in asymptotically AdS, dS and flat spacetimes for this black hole. In the presence of power-law Maxwell theory, $E(r)$ has a less divergence to the electrical field in linear Maxwell theory which by increasing the value of parameter $s$, this divergence becomes weaker. \\
In AdS spacetime and for small values of parameters $s$ and $q$, there is a black hole with two horizons where inner horizon is independent of the value of $M$ but the value of the outer horizon depends on the value of parameter $M$. Also, by increasing the value of $s$, there is just one horizon which increases as $M$ increases.\\
In this spacetime and for a constant value for $M$, there is a $s_{\max}$ which for $s<s_{max}$, there is a $q_{ext}$ which the solutions describe a black hole with two horizons for $q<q_{ext}$, an extreme black hole for $q=q_{ext}$ and a naked singularity for $q>q_{ext}$. Also for $s>s_{max}$, there is a $q_{ext2}$ which we have a naked singularity for $q>q_{ext2}$ and an extreme black hole for $q=q_{ext2}$. For $q<q_{ext2}$, there is a $q_{min}$ that we have a black hole with two horizons for $q_{min}<q<q_{ext2}$ and a nonextreme black hole for $q<q_{min}$. We should say that by increasing the value of parameter $s$, the values of $q_{ext}$ and $q_{ext2}$ decrease. \\
For constant parameters $s=1.1$, $q=1$, $M=4$, $\lambda=0.04$ and for $\mu<0$, if $|\mu|$ is so small, there is a black hole with two horizons, but for large $|\mu|$, we have a naked singularity. \\
dS and flat solutions are almost similar to AdS solutions. For example, for the same conditions and for $q=0.1$, general behaviors of dS and AdS solutions are similar to each other but the number of horizons in dS spacetime is one lower than the ones in Ads spacetime. \\
For constant values of parameters $M$ and $q$, the function $g(r)$ has a different behavior in $s=2$ than the other possible values of $s$. For example in dS spacetime, it is arranged that by increasing the value of $s$, the value of $r_{+}$ increases but the function $g(r)$ has a different behavior for $s=2$ and it has the least value for $r_{+}$ in $s=2$.  
For $s=2$ and constant values of $M$ in asymptotically dS spacetime, there is a $q_{1}$ which $g(r)$ is negative for all values of $r$ for $q<q_{1}$. But for $q>q_{1}$, $g(r)$ has a root in $r=r_{ext}$ which $g(r)$ is respectively negative and positive for $r>r_{ext}$ and $r<r_{ext}$. Also for the case $q>q_{1}$, by increasing the value of $q$, the value of $r_{ext}$ increases and for $r>r_{ext}$, the larger the value of $q$, the larger the value of $|g(r)|$.\\
We then obtained conserved and thermodynamic quantities and got to m-smarr formula by using them. All the obtained solutions satisfy the first law of thermodynamics. We also probed thermal stability for our solutions. We concluded that just AdS solutions of this black hole have thermal stability if the condition $T>0$ is established. \\
In this paper, we studied about the solutions of third order quasi-topological black hole in the presence of power-law Maxwell nonlinear electrodynamics. We thought that it may be also interesting to study solutions of fourth order quasi-topological gravity in the presence of the other nonlinear lagrangians like Logarithmic and exponential ones. So, we are doing this attractive subject. 

\acknowledgments{We would like to thank Payame Noor University and Jahrom
University.}

\end{document}